\documentclass[a4paper]{JHEP3}
\usepackage{epsfig}
\usepackage{graphicx}
\usepackage{amssymb}
\usepackage{latexsym}

\newcommand{\be}{\begin{equation}}
\newcommand{\ee}{\end{equation}}
\newcommand{\bea}{\begin{eqnarray}}
\newcommand{\eea}{\end{eqnarray}}

\newcommand{\tpd}[4]{{\left (\frac{\partial {#1}}{\partial {#2}}\right )^{#4}_{#3}}}

\def\dbar{{\mathchar22\mkern-12mu d}} 
\title{Does the first part of the second law also imply its second part?}
\author{N.D. Hari Dass
\\  Chennai Mathematical Institute, Chennai, India \\
CQIQC, Indian Institute of Science, Bangalore, India\\
Email: \email{dass@cmi.ac.in},\email{dass@cts.iisc.ernet.in}}

\abstract{Sommerfeld called the \emph{first part} of the second law to be the \emph{entropy axiom}, which is about the existence
of the state function \emph{entropy}. It was usually thought that the \emph{second part} of the second law, which is about the
non-decreasing nature of entropy of thermally isolated systems, did not follow from the first part. In this note, we point out
the surprise that the first part in fact implies the second part.
}

\keywords{Thermodynamics, Entropy, Second Law}

\begin{document}

\section{The Entropy Axiom: the first part of the second law}
Historically, the element of \emph{directionality} inherent in all thermal phenomena i.e that heat only flows from hotter
to colder bodies,
was stated most admirably through Clausius's formulation of the second law on the one hand, and
Kelvin's formulation on the other. Planck and Maxwell added their formulations which were essentially the same as these two,
but stated differently, laying emphasis on one aspect rather than the other. All of them were
expressions of \emph{the impossibility} of the so called \emph{perpetual machines of the second kind}. 

In this note,
we shall start with a very different approach; this is the perceptive remark by Sommerfeld \cite{sommerfeld}  that there
are two \emph{distinct} aspects of the second law. What he calls the first part is what can be called the \emph{entropy
axiom}, which postulates a state function that is now called \emph{entropy}. To the author's surprise, the first part
turns out to be fully equivalent to the entire second law! This will be explained in detail. As will be seen, the
entropy axiom is not some arbitrarily picked postulate; for ideal gases, this is in fact a \emph{consequence} of the
first law. 
\label{sec:entropyaxiom}
\section{A bonanza from first law for ideal gases}
As is well known, Joule's famous experiment on free expansion implies that for \emph{ideal
gases}, the internal energy U is a function of T only. In fact, this should be taken as an \emph{independent}
characterization of ideal gases from the gas law $PV=nRT$; it is independent because it can not be derived
from the gas laws without further assumptions. Now let us consider $\frac{\dbar Q}{T}$; according to 
the first law, considering only mechanical work,
\begin{equation}
\label{eq:Saxiomideal}
\frac{\dbar Q}{T} = \frac{dU(T)}{T}+nR\frac{dV}{V}=d{A(T)}+nRd\ln V
\end{equation}
where $U=\frac{dA(T)}{dT}$. This is a \emph{remarkable} result which states that for ideal gases, 
$\frac{dQ}{T}$
is indeed a perfect differential even though $\dbar Q$ was not.  This means that $\int\,\frac{\dbar Q}{T}$ is yet another state function. 
The interesting question is whether this is just an \emph{accident} valid only for ideal gases? Before attempting an
answer to it, let us call $\frac{\dbar Q}{T}$ as the perfect differential $dS$, where S shall be called \emph{entropy}. Clausius
introduced this concept in 1865, fifteen years after he formulated the first and second laws.
The entropy for the ideal gas, from above, is
\begin{equation}
\label{eq:idealS}
S = \int\,\frac{dU(T)}{T}+nR\ln V+S_0
\end{equation}
with $S_0$ being an \emph{undetermined constant}. When $C_V$ of the ideal gas is constant, this becomes
\begin{equation}
\label{eq:idealS2}
S = nC_V\ln T+nR\ln V+S_0\quad\quad S=nC_P\ln T-nR\ln P +S_0^\prime
\end{equation}

The existence of entropy as a state function is what Sommerfeld calls \emph{the entropy axiom}. We have seen that
for ideal gases this is not really an axiom, and is in fact a direct consequence of first law when combined with the
two laws for ideal gases. In general, U will not be a function of T alone and it is clear that the entropy axiom
will not always be valid.
\section{A consequence of the entropy axiom}
To explore the status of the entropy axiom for other than the ideal gases, let us 
consider gases obeying the van der Waals equation: 
\begin{equation}
\label{eq:vdW}
(P+\frac{an^2}{V^2})(V-nb) = nRT
\end{equation}
If we consider in particular the so called \emph{ideal} vdW gases, for which $C_V=const.$, it is not hard to see
that for the choice $U(V,T) = nC_VT-\frac{an^2}{V}$, $\frac{\dbar Q}{T}$ is again a perfect differential. This continues
to be so as long as $U(V,T) = f(T) -\frac{an^2}{V}$, but for choices other than these, $\frac{\dbar Q}{T}$ is indeed not
an exact differential.

Therefore what the entropy axiom does on one hand is restrict the possible choices of internal energy. In fact, one
can obtain a precise expression of this restriction by simultaneously demanding that $dU$ as well as $\frac{\dbar Q}{T}$
are perfect differentials. Taking (V,T) as the independent variables, these integrability conditions are, respectively,
\begin{equation}
\label{eq:dUintegra}
\tpd{C_V}{V}{T}{}=\left(\frac{\partial}{\partial T}\tpd{U}{V}{T}{}\right)_V
\end{equation}
and
\begin{equation}
\label{eq:dSintegra}
\left(\frac{\partial}{\partial V}\,\frac{1}{T}\tpd{U}{T}{V}{}\right )_T=\left (\frac{\partial}{\partial T}\,\frac{1}{T}
\left\{\tpd{U}{V}{T}{}+P\right\}\right )_V
\end{equation}
Simplifying this and using the previous equation, one arrives at one of the most important equations of thermodynamics i.e
\begin{equation}
\label{eq:dSintegra2}
\tpd{U}{V}{T}{}=T\tpd{P}{T}{V}{}-P
\end{equation}
Eqn.(\ref{eq:dSintegra2}) can be taken to be equivalent to the entropy axiom when the first law is valid.

When applied to an ideal gas for which $PV=nRT$, it is seen that this equation would require $\tpd{U}{V}{T}{}=0$
which is the same as U being a function of T alone. Earlier we had shown the converse i.e when U is a function of T
alone, the entropy axiom is satisfied. In the vdW case too, the particular form of $U(V,T)=f(T)-\frac{an^2}{V}$,
is indeed a solution of eqn.(\ref{eq:dSintegra2})!

The discussion of the entropy axiom so far seems rather mathematically oriented, without any obvious physical significance.
Actually, the entropy axiom has very deep physical significance, perhaps one of the deepest in physics! To bring this out, we
first demonstrate an \emph{equivalence} between the entropy axiom, and \emph{universality} of Carnot cycles.
\section{Entropy Axiom and Universality of Carnot Cycles}
\label{sec:entropyaxiomuniversality}
\begin{figure}[htbp]
\centering
\includegraphics[width=3.5in]{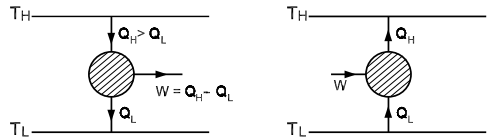}
\caption{The Carnot engine obeying first law and its reverse}
\label{fig:carnotnew}
\end{figure}
Within the caloric theory, Carnot had reached the very important conclusion that all ideal heat engines \emph{must}
have universal \emph{efficiency} if energy conservation (impossibility of perpetual mobiles of first kind, according to Carnot)
is to be respected. Now, with the new paradigm of interconvertability of heat and work, it is pertinent to raise afresh the issue
of the universality of all ideal Carnot cycles. 

But unlike in the caloric theory, now the Carnot cycle is characterized by the amount of heat $Q_H$ absorbed by the 
system at the higher temperature $T_H$, and the heat $Q_L < Q_H$ relinquished at the lower temperature $T_L$.
Even in the new theory, the notion of efficiency can still be kept to mean
the amount of work performed per heat absorbed at the higher temperature. 

The important difference from the caloric theory
is that the heat relinquished at the lower temperature is no longer the same as that absorbed at the higher temperature, but is
in fact reduced by the amount of work performed. Consequently, the efficiency is given by $e = \frac{\Delta W}{Q_H}=\frac{Q_H-Q_L}{Q_H}$.
What is not clear a priori is that in the new theory, the ratio $\frac{Q_L}{Q_H}$ is universal for all ideal heat engines. 

What, if any, would go wrong if the efficiencies of all ideal heat engines were not the same? Precisely the same kind
of analysis that Carnot carried out earlier can be done now too. The ideal Carnot cycle and its reverse are shown in
the next figure. In the reverse engine, heat $Q_L$ is absorbed at the lower temperature and $Q_H > Q_L$ is exhausted
at the higher temperature after work equal to $Q_H-Q_L= eQ_H$ has been performed \emph{on} the system. If there were
two ideal Carnot engines $C,C^\prime$ with efficiencies $e, e^\prime > e$, then a composite of $C^\prime$ with the 
reverse of C (see figure) would extract an amount of heat $(e^\prime-e)Q_H$ from the lower reservoir and convert it 
\emph{completely} to work W of same magnitude, without expelling \emph{any} heat at the higher reservoir. In other
words, \emph{a perpetuum mobile of the second kind} would be possible! This can only be avoided if the efficiencies
of all ideal Carnot cycles, even in the new theory, are the same. 

\begin{figure}[htbp]
\begin{minipage}[b]{0.5\linewidth}
\centering
\includegraphics[width=\linewidth]{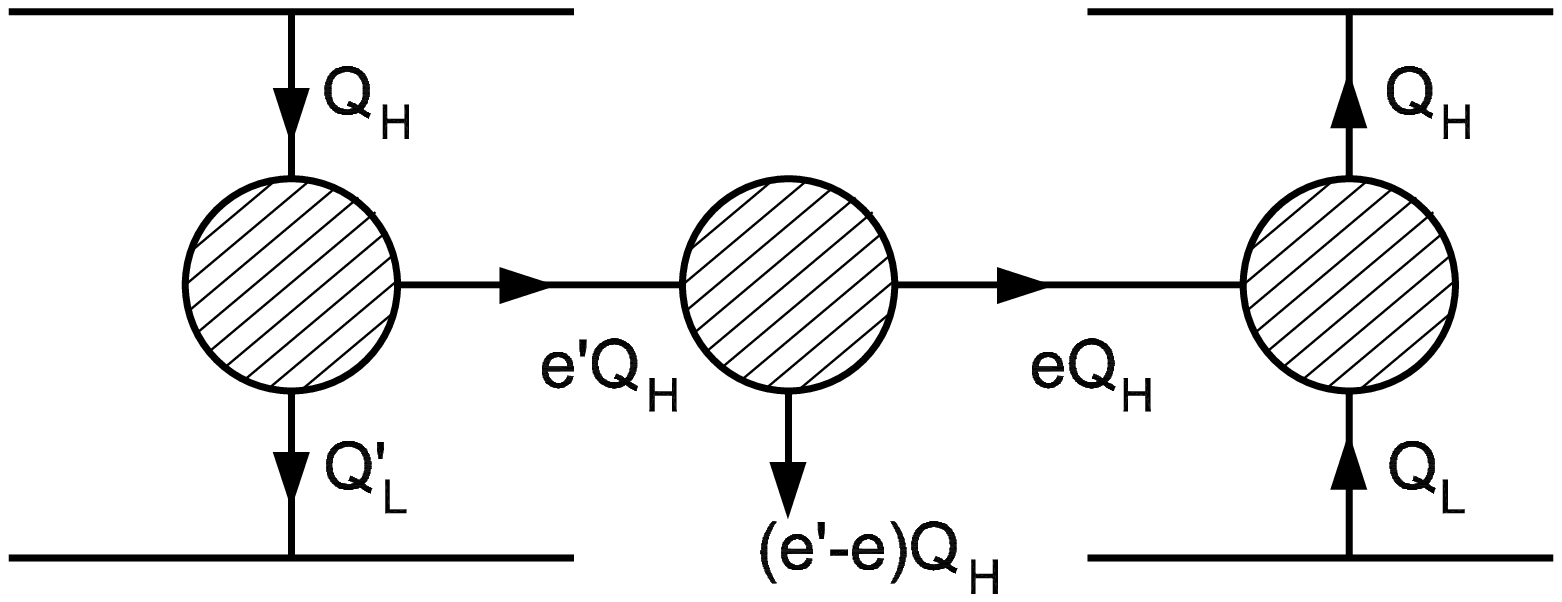}
\caption{Universality of the Carnot cycles}
\label{fig:carnotuniversal}
\end{minipage}
\hspace{0.5 cm}
\begin{minipage}[b]{0.5\linewidth}
\centering
\includegraphics[width=\linewidth]{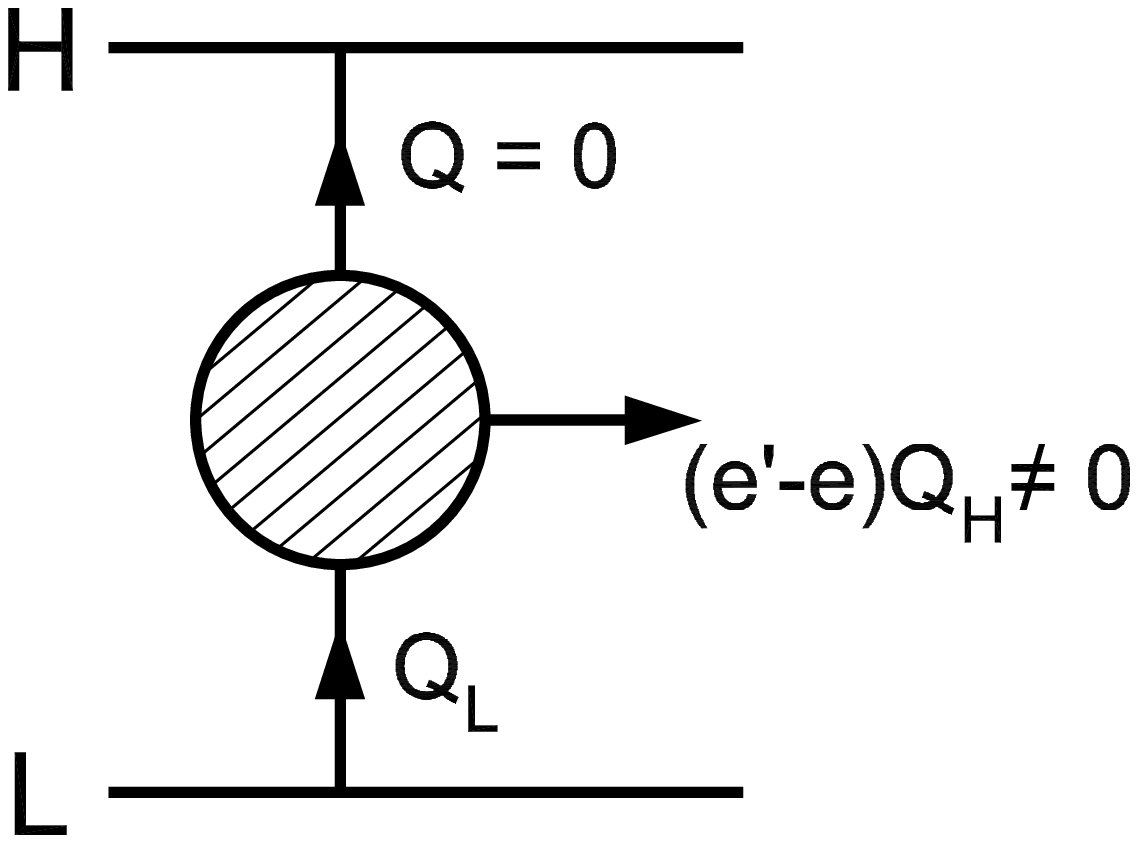}
\caption{Universality of the Carnot cycles}
\label{fig:carnotuniversal2}
\end{minipage}
\end{figure}
This line of thinking, as carnot had already demonstrated earlier, puts severe restrictions on efficiencies of
\emph{nonideal}(i.e not reversible) heat engines too. 
While the reverese of an irreversible(nonideal) Carnot cycle cannot be used, the argument can still be made use
of with the irreversible Carnot cycle in combination with the reverse of a reversible cycle; therefore, if the
efficiency of an irreversible cycle \emph{exceeds} that of an ideal Carnot cycle, one can still construct a perpetual
mobile of the second kind. Therefore one can conclude that \emph{efficiency of an irreversible engine has to be necessarily lower 
than a reversible engine}.

We will now proceed to determine the conditions for the universality of ideal engines.
We can just use the same calculus that Clausius used to establish the existence of internal energy U. So, considering an
\emph{infinitesimal Carnot cycle}, the efficiency de is given by $de=\frac{\Delta W}{N(V,T)\delta_T V}$. On using the results
previously obtained, one gets $de=\frac{1}{N(V,T)}\tpd{P}{T}{V}{}\,\delta T$.
Clausius is supposed to have been aware that for ideal gases $N(V,T)=P$. This is also what follows from Joule's experiment 
on expansions of ideal gases. In the light of the first law, this implies that for ideal gases U is a function of T only,
and consequently $dU(T)=C_V(T)dT$. On combining this with eqn(2.11) one concludes that $N=P$. Therefore, for ideal
gases 
\begin{equation}
\label{eq:idealcarnotdeff}
de= \frac{1}{P}\tpd{P}{T}{V}{}dT = \frac{dT}{T} 
\end{equation}
This is how Clausius fixed the universal Carnot function in
the earlier Clapeyron equation to be $\mu(T) = \frac{1}{T}$ leading to the modern form of the Clapeyron equation which
should aptly be called the \emph{Clausius-Clapeyron equation}. However, to avoid confusion because of this equation being referred
to in current literature as \emph{Clapeyron equation}, we shall also continue with that practice with the understanding that the latter
is a shorthand for the former. It is of course to be recalled that the earlier Clapeyron
equation was based on the now defunct caloric theory, whereas Clausius derivation is what follows from the new theory of heat;
it is just that many quantities(but of course not all) were insensitive to the actual nature of heat!

Now the requirement of universality of efficiencies of all ideal heat engines means that in particular they must equal the
efficiency of all ideal heat engines based on ideal gases as the working substance, and one gets the extremely important
consequence that $N(V,T)=T\tpd{P}{T}{V}{}$ \emph{for all thermodynamic systems}! On the other hand, $\tpd{U}{V}{T}{}=N-P$.
Therefore, universality of efficiencies of Carnot engines, in the light of the first law, requires that
\begin{equation}
\label{eq:Cuniversality}
\tpd{U}{V}{T}{}=T\tpd{P}{T}{V}{}-P
\end{equation}
Lo and behold, this is nothing but the condition for entropy axiom! In other words, \emph{the entropy axiom is equivalent
to the condition of universality of all Carnot engines}. This is the underlying physical significance of the entropy
axiom. But since the universality is also equivalent to the impossibility of perpetual mobiles of the second kind, we draw
the powerful conclusion that the entropy axiom, what Sommerfeld called the first part of the second law, is equivalent to
the impossibility of perpetual mobiles of second kind. But the latter is one of the formulations of the \emph{second law
of thermodynamics}, what Sommerfeld would have called the \emph{second part of the second law}. Therefore, the entropy
axiom is not just the first part of the second law, it is, at the same time, also its second part!

However, most people recognize second law in the form where it states that \emph{entropy of a thermally isolated system
never decreases}. As the system while executing a Carnot cycle is certainly not \emph{thermally isolated}(except during
the adiabatic stages), one can not immediately see the consequence of the second law, so formulated, for a Carnot cycle.
Instead, we shall focus on the so called \emph{Clausius inequality} which we shall later see to be equivalent to the form
of the second law so stated. This inequality states that for \emph{any} cycle, not necessarily reversible, the following is
always true:
\begin{equation}
\label{eq:clsinequality}
\oint \frac{\dbar Q}{T} \le 0
\end{equation}
The equality holding only for reversible cycles. It is very important to stress that though $\frac{\dbar Q}{T}$ is a perfect differential, the integral of this over a path
is the entropy difference between the endpoints only if the path is reversible. That is why even though $\frac{\dbar Q}{T}$
has been integrated over a closed path in eqn.(\ref{eq:clsinequality}), the rhs is not zero! Furthermore,
the rhs can be \emph{negative} so a naive interpretation of the integral as a change of entropy would actually imply
an entropy decrease!  We shall return to a fuller discussion of these subtleties shortly.

We shall now
demonstrate that the entropy axiom, through its equivalence to the impossibility of perpetual mobiles of the second kind via
its equivalence to the universality of ideal Carnot cycles, indeed yields
the Clausius inequality, \emph{without any further assumptions}. For that we make use of Clausius's own ingeneous
construction.

Consider an \emph{arbitrary} cycle ${\tilde C}$, not necessarily a reversible ones. In executing this, let the system start
at $A_1$, and absorb an amount of heat $(\Delta Q)_1$ during the segment $A_1A_2$ at temperature $T_1$. The cycle is completed
by absorbing $(\Delta Q)_2$ during $A_2A_3$ at $T_2$, and so on, till the system returns to its starting state $A_1$ by
absorbing $(\Delta Q)_n$ during $A_nA_1$ at $T_n$. The sign of $(\Delta Q)$ can be positive or negative.

\begin{figure}[htbp]
\centering
\includegraphics[width=1.5in]{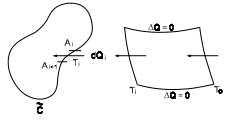}
\caption{Proving the Clausius inequality.}
\label{fig:clausineq}
\end{figure}
Clausius's ingenuity lay in picturing the heat absorbed at each stage as the heat relinquished during a \emph{reversible}
Carnot cycle operating between some arbitrary temperature $T_0$ and the temperature of the stage of ${\tilde C}$ during
which the heat was absorbed. Clearly, there are, in addition to the cycle ${\tilde C}$, n Carnot cycles $C_1,C_2,..,C_n$
operating between $T_0$ and the temperatures $T_1,T_2,..,T_n$. This is schematically shown in the figure for the stage
$A_iA_{i+1}$. 

So, the cycle ${\tilde C}$ starts at $A_1$ and at the same time $C_1$ starts at $E_1$ and goes through the reversible
Carnot cycle by eventually delivering $(\Delta Q)_1$ to ${\cal C}$ during $A_1A_2$ at temperature $T_1$. It must therefore 
absorb $\frac{T_0}{T_1}(\Delta Q)_1$ from the reservoir at $T_0$. After ${\tilde C}$ and $C_1,..,C_n$ have been completed,
all have returned to their original states; the total heat absorbed from the \emph{single} reservoir at $T_0$ being
\begin{equation}
\label{eq:clsheatsum}
\Delta Q = T_0\sum_i\,\frac{(\Delta Q)_i}{T_i}
\end{equation}
and this is completely converted to work, with no other changes. If this heat were \emph{positive}, we would indeed have 
realized a perpetual machine of the second kind. On the other hand, if this heat were \emph{negative}, the work done would have been
\emph{on} the system, and this would only have amounted to a \emph{refrigerator}, with no contradictions. The case when $\Delta Q=0$
does not also contradict anything.

In conclusion, the entropy axiom is completely equivalent to a) universality of all ideal heat engines, b) the impossibility
of perpetual machines of second kind, and consequently, c) Clausius inequality of eqn.(\ref{eq:clsinequality}). 
The standard form of the second law stating that entropy of thermally isolated systems never decreases emerges from
the Clausius inequality, as expounded in several books on Thermodynamics..

\acknowledgments
The author acknowledges support from the Department of Science and Technology to the project IR/S2/PU-001/2008.


\begin{thebibliography}{}
\bibitem{sommerfeld} 
A. Sommerfeld, \it{Thermodynamics and Statistical Mechanics}, Levant Books, Kolkata 2005(Reprinted from the original).
\end{thebibliography}
\end{document}